\documentclass[namedreferences]{solarphysics}
\usepackage[optionalrh,natbib]{spr-sola-addons} 
\usepackage{graphicx}        
\usepackage{amssymb}        
\usepackage{color}           
\usepackage[T1]{fontenc}
\usepackage{url}             

\usepackage[psamsfonts]{amsfonts}
\usepackage{mathptmx}
\usepackage{booktabs}

\definecolor{darkgreen}{RGB}{0,142,128}

\usepackage[pdfborder={0 0 0},urlcolor=blue,breaklinks,colorlinks,linkcolor=blue,citecolor=blue]{hyperref}
\usepackage{breakurl}

\newcommand{\aap}{    {\it Astron. Astrophys.}}

\newcommand{\apj}{    {\it Astrophys. J.}}
\newcommand{\apjl}{   {\it Astrophys. J. Lett.}}

\newcommand{\lrsp}{  {\it Living Rev. Solar Phys.}}

\newcommand{\pra}{    {\it Phys. Rev. A}}

\newcommand{\prl}{    {\it Phys. Rev. Lett.}}
\newcommand{\solphys}{{\it Solar Phys.}}

\newcommand{\changed}[1]{{#1}}
\newcommand{\Paulchanged}[1]{{#1}}

\newcommand{\AStwo}[1]{{#1}}

\newcommand{\ASthree}[1]{{#1}}

\begin{document}

\begin{article}

\begin{opening}

\title{Predictive Capabilities of Avalanche Models for Solar Flares}
\author{A. Strugarek, P. Charbonneau}
\institute{D\'epartement de physique, Universit\'e de Montr\'eal, C.P. 6128 Succ. Centre-Ville,
Montr\'eal, Qc, H3C-3J7, CANADA}
\date{\today}

\begin{abstract}

We assess the predictive capabilities of various classes of avalanche
models for solar flares. We demonstrate that avalanche models cannot
generally be used to predict specific events due to their high
sensitivity to their embedded stochastic process. We show that
deterministically driven models can nevertheless alleviate this caveat
and be efficiently used for large events predictions. Our results
promote a new approach for large (typically X-class) solar flares
predictions based on simple and computationally inexpensive avalanche models. 

\end{abstract}

\keywords{Flares, forecasting; Avalanche models; Self-organized criticality}

\end{opening}

\section{Introduction}
\label{sec:introduction}

It has now been clearly established that solar eruptive phenomena have
multiple incidences on the heliosphere, and in particular on the Earth
space environment. Along with coronal mass ejections
\citep[CMEs, see][]{Chen:2011gq}, solar flares are one of the most dangerous space
weather events. They are also systematically observed before a large
CME is triggered \citep{Shibata:2011kd}. While CME-triggered energetic
particles reach Earth on time scales of tens of hours, high energy
protons accelerated by a flaring process may reach 1 AU few tens of minutes
later, and the associated X-ray photons affecting the Earth's ionosphere only 8 minutes
later. No robust precursor of solar flares have been identified so far, preventing any
efficient empirical forecast. An ongoing significant
effort to predict when a flare of a given magnitude will occur have been pursued over
the last decade \citep[for a recent review, see][and references
therein]{Georgoulis:2012jr}, with moderate success so far. 
Any progress in this direction is thus highly valuable for risk management
related to space weather.

Decades of flare observations have shown that the probability
distribution function $f(E)$ for flare energy $E$ takes the form of a
power law, $f(E)\propto E^{-\alpha}$ spanning some 8 orders of
magnitude in energy, with estimated in the range $1.4-2.0$
\citep{Aschwanden:2002jy,Aschwanden:2011hw}. 
Avalanches provide one class of physical phenomena characterized by
such scale-free energy release when driven slowly and
continuously. Many avalanche models have been developed to represent
the solar flares distribution \citep[see][and references
therein]{Charbonneau:2013vt}, most of which are based on the concept of
self-organized criticality \citep[hereafter SOC;
see][]{Bak:1987ef,Jensen:1998ww}. \citet[][hereafter LH]{Lu:1991hp} have proposed an
avalanche-type model for solar flares that has become a
reference model in the solar context, although numerous variations
have now been developed over the intervening years. 


If flares are indeed accurately described by an avalanche-type model,
forecasting them may appear dubious because of the stochastic driving
from which they originate. In a lattice-based avalanche model -- such
as the LH model --, the size of an avalanche is controlled by
(i) the starting position of the instability triggering it and (ii) the closeness to
the threshold of the neighbor nodes. In the original LH model,
the starting position mainly depends on the stochastic driver while
the state of the system is the complex result of past avalanching
history. If a large portion of the system is close to the threshold,
the short term avalanching behavior may depend marginally on the
stochasticity of the driver. Building on this idea,
\citet{Belanger:2007ey} coupled the LH model to data assimilation
technique to forecast solar flares. It remains unclear, though, whether a
mean prediction (\textit{i.e.}, when varying the stochastic driver) can
generally be defined from such an avalanche model. 

\changed{This
paper focuses on the following question: can avalanches models
be used for
predictive purposes in the context of solar flares? We present a study
of the predictive capabilities of a series of
avalanches models, starting from the original LH model, in which we introduce
variations in the stochastic components -- the latter, as expected, determining the
predictive capability of the model.}
The models considered and their statistical
properties are described in section \ref{sec:aval-models-solar} and
\ref{sec:model-properties}. Section \ref{sec:predictive-abilities}
provides \changed{a methodology} to define a prediction from an
avalanche model \changed{as well as} estimates
of the predictive capabilities of the models we considered. We conclude in
section \ref{sec:conclusions} by a discussion summarizing our results
\Paulchanged{and the identification of subclasses of avalanche models most
promising}
for the development of a
solar flares prediction tool based on avalanche models.

\section{Avalanche Models for Solar Flares}
\label{sec:aval-models-solar}

An avalanche model necessarily includes some kind of stochasticity. We
develop different SOC models by modifying how and where the stochasticity
appears, which will lead to very different predictive capabilities
(see section \ref{sec:predictive-abilities}).
\Paulchanged{With one exception
(see section \ref{sec:georgoulis--vlahos})},
the models we use
are described in details in \citet{Strugarek:2014kj}. We summarize briefly their
properties and defer the interested reader to this other paper.

\subsection{The Lu and Hamilton Model}
\label{sec:lu--hamilton}

Following in part \citet{Kadanoff:1989ki}, \citet{Lu:1991hp} have
developed a SOC avalanche model for solar flares that by now has
become a kind of ``standard'' 
\citep[for a review, see][]{Charbonneau:2001fv}. Here we consider a
version of the LH model defined over a 2D regular cartesian grid with
nearest-neighbor connectivity over which a scalar
field $A^{n}_{i,j}$ is defined. The superscript $n$ is a discrete time
index, and the subscript pair 
$(i,j)$ identifies a single node on the 2D lattice. Keeping $A=0$ on the lattice
boundaries, the cellular automaton is driven by adding one small
increment in $\delta A$ per time step, at some randomly selected node that
changes from one time step to the next. A deterministic stability
criterion is defined in terms of the local curvature of the field at
node $(i,j)$: 
\begin{equation}
  \label{eq:delta_A_LH}
  \Delta A^n_{i,j} \equiv A^n_{i,j} - \frac{1}{4}\sum_k A^n_k
  \, ,
\end{equation}
where the sum runs over the four nearest neighbors at nodes $(i,j\pm
1)$ and $(i\pm 1, j)$. If this quantity exceeds some preset threshold
$Z_c$ then an amount of nodal variable $Z$ is redistributed to the same
four nearest neighbors according to the following discrete,
deterministic rules: 
\begin{eqnarray}
  \label{eq:redistribution1_LH}
  A^{n+1}_{i,j} &=& A^{n}_{i,j}  -\frac{4}{5}Z\, , \\
  \label{eq:redistribution2_LH}
  A^{n+1}_{i\pm 1,j\pm 1} &=& A^{n}_{i\pm 1,j\pm 1}  +\frac{1}{5}Z\, ,
\end{eqnarray}
where $Z\equiv Z_c\, \Delta A^{n}_{i,j}/\left|\Delta A^{n}_{i,j}\right|$. Following this 
redistribution it is possible that 
one of the nearest-neighbor nodes now exceeds the stability
threshold. The redistribution process begins anew from this node, and
so on in classical avalanching manner. Driving is suspended during
avalanching, implicitly implying a separation of timescales between
driving and avalanching dynamics, and all nodal values are updated
synchronously during avalanche to avoid introducing a directional bias
in avalanche propagation. 

It is readily shown that these redistribution rules, while
conservative in $A$, lead to a decrease in $A^2$ summed over the five
nodes involved by an amount: 
\begin{equation}
  \label{eq:incr_energy_LH}
  \Delta 
e^n_{i,j} = \frac{4}{5}\left(2\frac{\left|\Delta A^n_{i,j}\right|}{Z_c} -1
  \right)Z_c^2\, ,
\end{equation}
with the energy released being ``assigned'' to the unstable node
$(i,j)$. If one identifies $A^2$ with a measure of magnetic energy
\citep[see][]{Charbonneau:2013vt},
the total energy liberated by all unstable
nodes at a given iteration is then equated to the energy release per
unit time in the flare.
A natural energy unit here (used for normalization in all that
follows), is the quantity 
of energy $e_0\equiv 4Z_{c}^{2}/5$ liberated by a single node exceeding the stability
threshold by an infinitesimal amount. 
This very simple model yields a good representation of
flare statistics, namely the observed power-law form (and associated
exponents) of the frequency distributions of flare peak energy release
$P$, duration $T$, and total energy release $E$
\citep{Lu:1993cy,Charbonneau:2001fv,Aschwanden:2002fl}.

\subsection{The Georgoulis and Vlahos Model}
\label{sec:georgoulis--vlahos}

\citet[][hereafter GV]{Georgoulis:1996bs,Georgoulis:1998tp} have developed a variation
of the LH model based on the work of \citet{Vlahos:1995vu}
by adding anisotropic stability criterion and redistribution
rules. This model is known to 
generate a double power-law frequency  
distribution of avalanche parameters E, P and T, with a steeper
power law for the smaller events \citep{Georgoulis:1996bs}. 

Then, they introduced a 
modified driving scheme \citep{Georgoulis:1998tp} using a 
power-low random number generator for the small 
increments of $\delta A$. The increments follow the probability distribution 
\begin{equation}
  \label{eq:GV_powerlaw}
  P(\delta A) \propto (\delta A)^{-\alpha}\, .
\end{equation}
They showed that the power law indexes of the avalanche parameters E, P
and T vary roughly linearly with the parameter $\alpha$. This method
provides a very interesting mean of controlling the power-law indexes
of SOC models. We choose, for the present study, to retain only this
modified driving scheme -- leaving out the anisotropic stability
criterion and redistribution rules -- from the complete original GV model.

In addition to the GV model, \citet{Norman:2001co} also developed a modified
driving scheme by rendering it non-stationary. This work was motivated initially
by the results of \citet{Wheatland:2000if}, who showed that the waiting time
distribution of solar flares should be characterized by a power-law
tail for long waiting times,
rather than by an exponential decay which is obtained from the LH
model. Finally, \citet{Aschwanden:2010eq} re-analyzed the solar flare data from various
source and concluded that the waiting time distribution of solar
flares is indeed consistent with a non-stationary Poisson process, and
can well be described with an avalanche model possessing a
non-stationary driver. 

A biased driving scheme could in principle make the model either more
or less robust with respect to various random number sequences, and
hence improve or decrease its
predictive capabilities. In order to simplify the discussion we only
detail the results obtained with the GV model in this work. Similar
analysis made with the model of \citet{Norman:2001co} lead to the
same conclusions regarding its predictive capabilities (not shown
here). In the following we use a GV model with a parameter $\alpha=2.6$.

\subsection{The Deterministically Driven Model}
\label{sec:deterministic-model}

The random driver of the LH and GV models can be nicely linked to the 
Parker picture of random shuffling of a loop's magnetic footpoint
by photospheric flows. For loop diameters smaller than this scale,
though, the granular flow 
displaces the footpoints in a spatially coherent manner far removed
from random shuffling. One particularly interesting form of such
global forcing is a twisting of the loop's footpoints, which then
propagates upwards and accumulates along the length of the loop. This
form of global forcing has a direct equivalent in our 2D cellular
automaton \citep[][]{Strugarek:2014kj} 
\Paulchanged{through the driving rule}
\begin{equation}
  \label{eq:drive_deter}
  A^{n+1}_{i,j} = A^n_{i,j}\times\left(1+\varepsilon\right)\, , \,\,\,
  \varepsilon \ll 1\, , \,\,\, \forall\left( i,j \right)\, ,
\end{equation}
where the parameter $\varepsilon$ ($\ll 1$) is a measure of the
driving rate. As in the LH model, driving is interrupted during
avalanching, which amounts to assuming that the driving timescale is
much longer than the avalanching timescale, a reasonable assumption in
the solar coronal context \citep[for a discussion, see][]{Lu:1995ce}.



The stochastic component of the deterministically driven model can appear in
the threshold definition, in the extraction or in the redistribution rule. In all rules
discussed so far, redistribution is conservative, in that whatever 
quantity of $A$ being extracted from an unstable node ends up in the
 nearest neighbors. This conservation property is basically
 inspired by the sandpile analogy, where avalanches redistribute sand
 grains without creating or destroying any. 
 A nonconservative version of the LH redistribution rules can be defined as follows:
\begin{eqnarray}
  \label{eq:redistribution1_D}
  A^{n+1}_{i,j} &=& A^{n}_{i,j}  -\frac{4}{5}Z\, , \\
  \label{eq:redistribution2_D}
  A^{n+1}_{i\pm 1,j\pm 1} &=& A^{n}_{i\pm 1,j\pm 1}  +\frac{r_{0}}{5}Z\, ,
\end{eqnarray}
where $r_{0}\in [D_{nc},1]$ is again extracted from a uniform distribution
of random deviates with a lower bound $D_{nc}\, (<1)$, such as $1-D_{nc}$  is
the fraction of the redistributed quantity $Z$ that is lost rather
than redistributed. This rule thus involves one free parameter, namely
the conservation parameter $D_{nc}\in]0,1[$. A nonconservative model of this
type, using fully deterministic driving, and random redistribution and stability
criteria, has been studied extensively in \citet{Strugarek:2014kj}. We consider
here \AStwo{three deterministically driven models (hereafter D1, D2
  and D3, or D models) with different levels of stochasticity. Model D1 corresponds
to model NC6 in \citet{Strugarek:2014kj} and includes random
extraction, random redistribution and random non-conservation
components. Model D2 corresponds to model NC0, which involves only one
type of stochastic process located in the non-conservative
redistribution rule ($D_{nc}=0.1$). Finally, model D3 is equivalent to D2 with a
significantly lower non-conservation degree ($D_{nc}=0.9$)}.

\section{Models Properties}
\label{sec:model-properties}

\changed{In the following analysis, we test the predictive
  capabilities of the \AStwo{five} models by combining $2000$ different random
  number sequences for $200$ different initial conditions for a total
  of $2\, 10^{6}$ runs of (on average) $10^{4}$ iterations. The statistical
  properties of each model (described hereafter) are obtained from longer individual
  runs of more than $10^{7}$ iterations. The large number of runs we
  considered is necessary for assessing the predictive skills of each
  model but precludes the use of large lattices. We run the \AStwo{five}
  models on a mid-size [$48\times48$] lattice which represents a good
  computational compromise since the power-law exponents
  and global properties of the models were shown to vary only
  marginally for larger lattices in the considered models \citep{Vlahos:1995vu,Charbonneau:2001fv,Strugarek:2014kj}.}

\begin{figure}
  \centering
  \includegraphics[width=\linewidth]{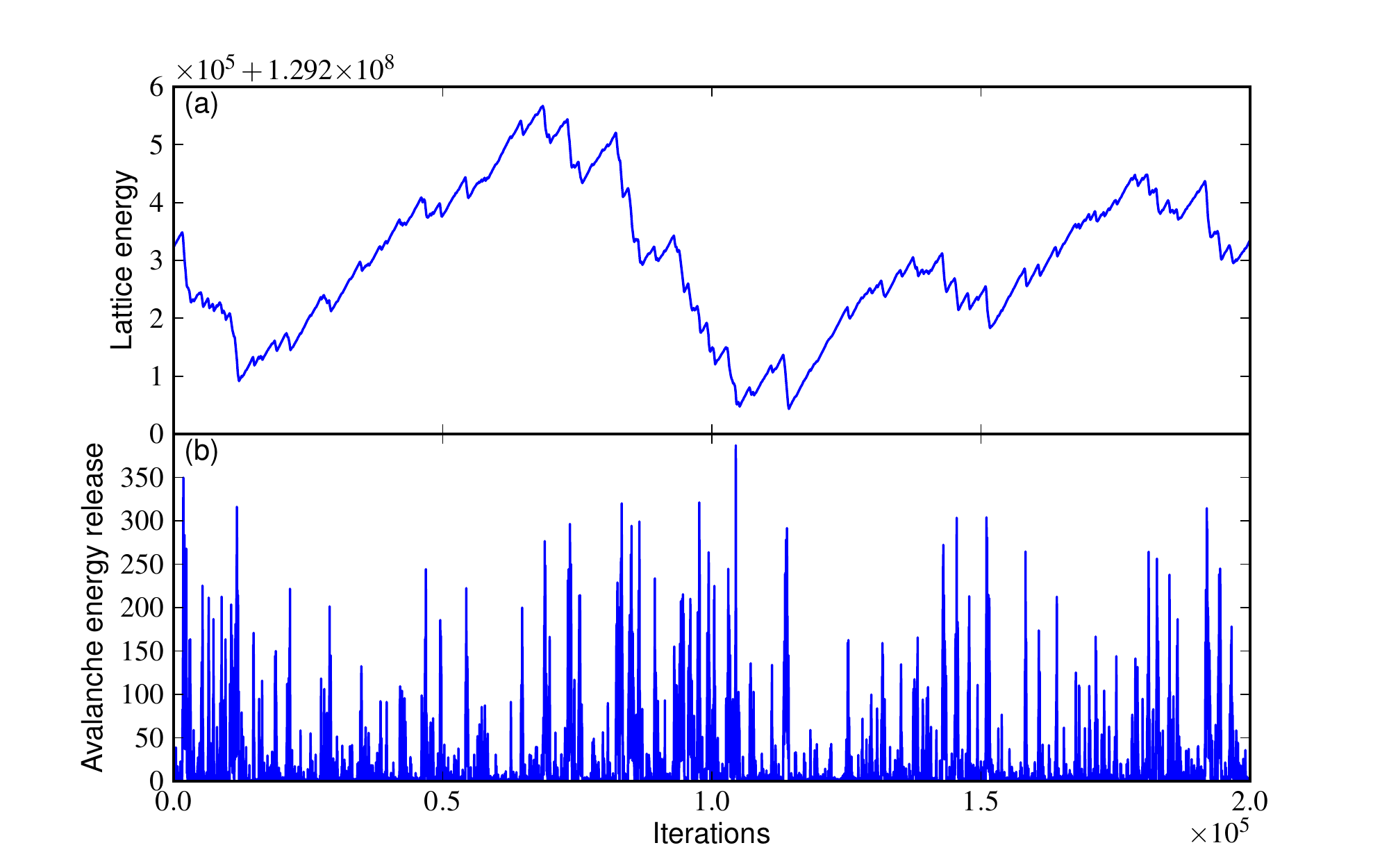}
  \caption{Lattice energy (a) and avalanche energy release (b) time
    series for the LH model. Both energies are normalized to $e_{0}$.}
  \label{fig:lh_example}
\end{figure}

Fig. ~\ref{fig:lh_example} shows a small sample of a time series for
the lattice energy (a) and avalanche energy release (b) in
the LH model. The lattice energy fluctuates around a mean state and
energy is released by avalanches of various duration and size.

\begin{figure}
  \centering
  \includegraphics[width=\linewidth]{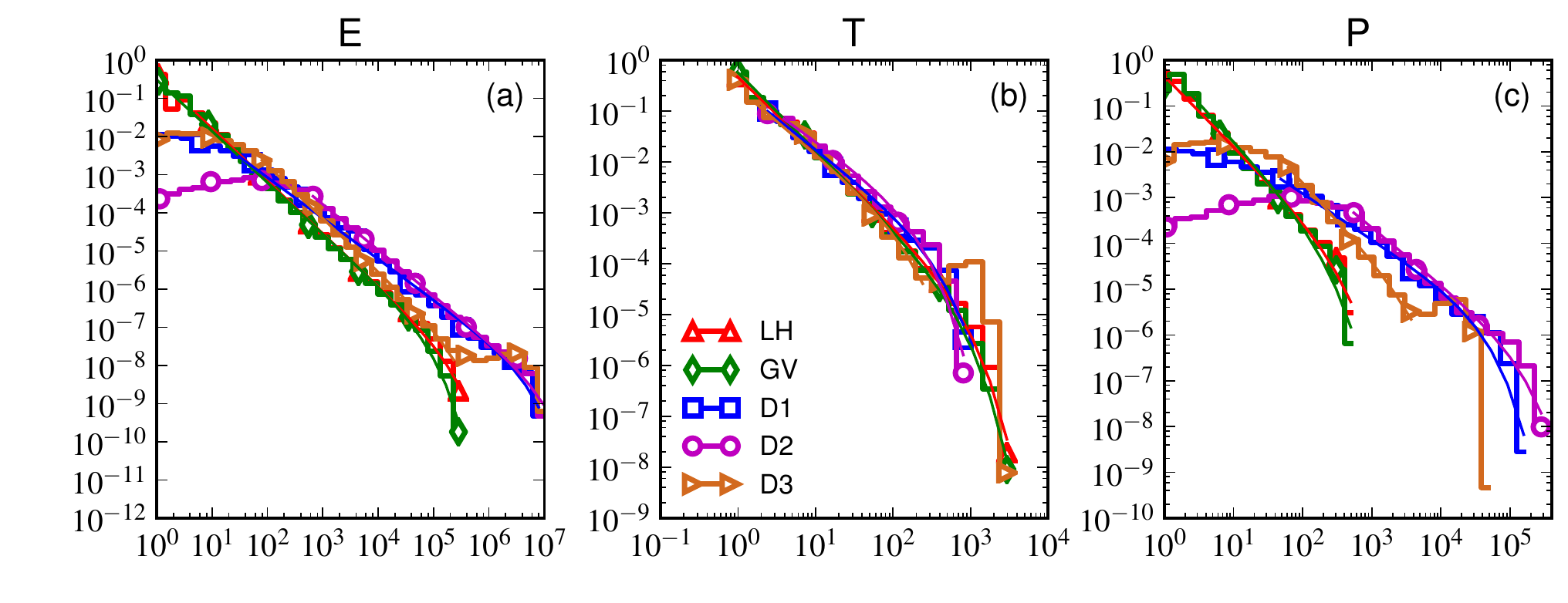}
  \caption{Probability density functions of avalanches properties for the different models. The energy E
  (a), duration T (b) and peak energy P (c) of avalanches are
  shown. The power-law exponent for each quantity and each model are
  indicated in table \ref{tab:props_mods} and the fits are plotted (solid lines).}
  \label{fig:mod_properties}
\end{figure}

\AStwo{The total energy release (E), duration (T) and peak energy (P) of
avalanches are distributed as a power-law over several decades (see
the probability distribution functions -- PDFs -- in
fig. \ref{fig:mod_properties}). The D models have an anomalous
non-power law component for very small avalanches, and model D3
  also exhibit and small excess of large avalanches
\citep[for a complete discussion on these models,
see][]{Strugarek:2014kj}. In this work we assess the capability
of avalanche models to predict large avalanche. 
Hence, the peculiar small avalanches
population does not affect the results of this paper. 
The GV model possess a significantly different waiting time
statistics (not shown here). Because of all these differences, we need 
to precisely define the time intervals and 
avalanches properties on which we characterize the predictive
capabilities to ensure that some specificities of the models do not
include some bias to our analysis. These definitions will naturally change from one model
to the other (see section \ref{sec:predictive-abilities}).}

\AStwo{The five models 
\Paulchanged{are characterized by} different power-law exponents, which are listed
in table \ref{tab:props_mods}. The power-law exponents
characterizing the LH, GV and D3 models
stand at the very low end of the observationally-inferred value for
solar flares, \
citep[e.g., $\alpha_{E} \in [1.39,1.78 {]}$ in the hard X-rays data, see][]{Aschwanden:2014wg},
and fall significantly below this range for the D1 and D2 models.
We will see that the predictive capabilities of the avalanches models
are rather insensitive to these power-law exponents, but rather strongly
depend on their embedded stochasticity 
Note that it is in principle possible to alter the components of the model to
better reproduce the observationally-inferred power-law indices
\citep[see][]{Charbonneau:2001fv,Strugarek:2014kj,Aschwanden:2014wg}.}

\begin{table}
\centering
\begin{tabular}{lrrrrr}
\toprule
{} &     $E_0$ & $\tau_w$ & $\alpha_E$ & $\alpha_P$ & $\alpha_T$ \\
\midrule
LH &   2 10$^4$ &   3.25 10$^{3}$ &     $1.37$ &    $1.53$ &     $1.45$ \\
GV &   2 10$^4$ &   1.5 10$^{3}$  &     $1.35$ &     $1.66$ &     $1.53$ \\
D1 &  2 10$^6$ &    10$^{3}$ &  $1.07$ &     $1.00$ &     $1.21$ \\
D2 &  2 10$^6$ &   1.5 10$^{3}$ &  $1.20$ &     $1.24$ &     $1.08$ \\
D3 &  2 10$^6$ &   1.5 10$^{3}$ & $1.39$ &     $1.63$ &     $1.28$ \\
\bottomrule
\end{tabular}
\caption{Statistical properties of the \AStwo{five} models
  used for this study. The normalization energy and time windows are
  also indicated (see section \ref{sec:predictive-abilities}).
\Paulchanged{All models are defined on a 2D $48\times 48$ cartesian lattice
with four-nearest-neighbour connectivity.}}
\label{tab:props_mods}
\end{table}

\section{Predictive Capabilities}
\label{sec:predictive-abilities}

\begin{figure}
  \centering
  \includegraphics[width=0.7\linewidth]{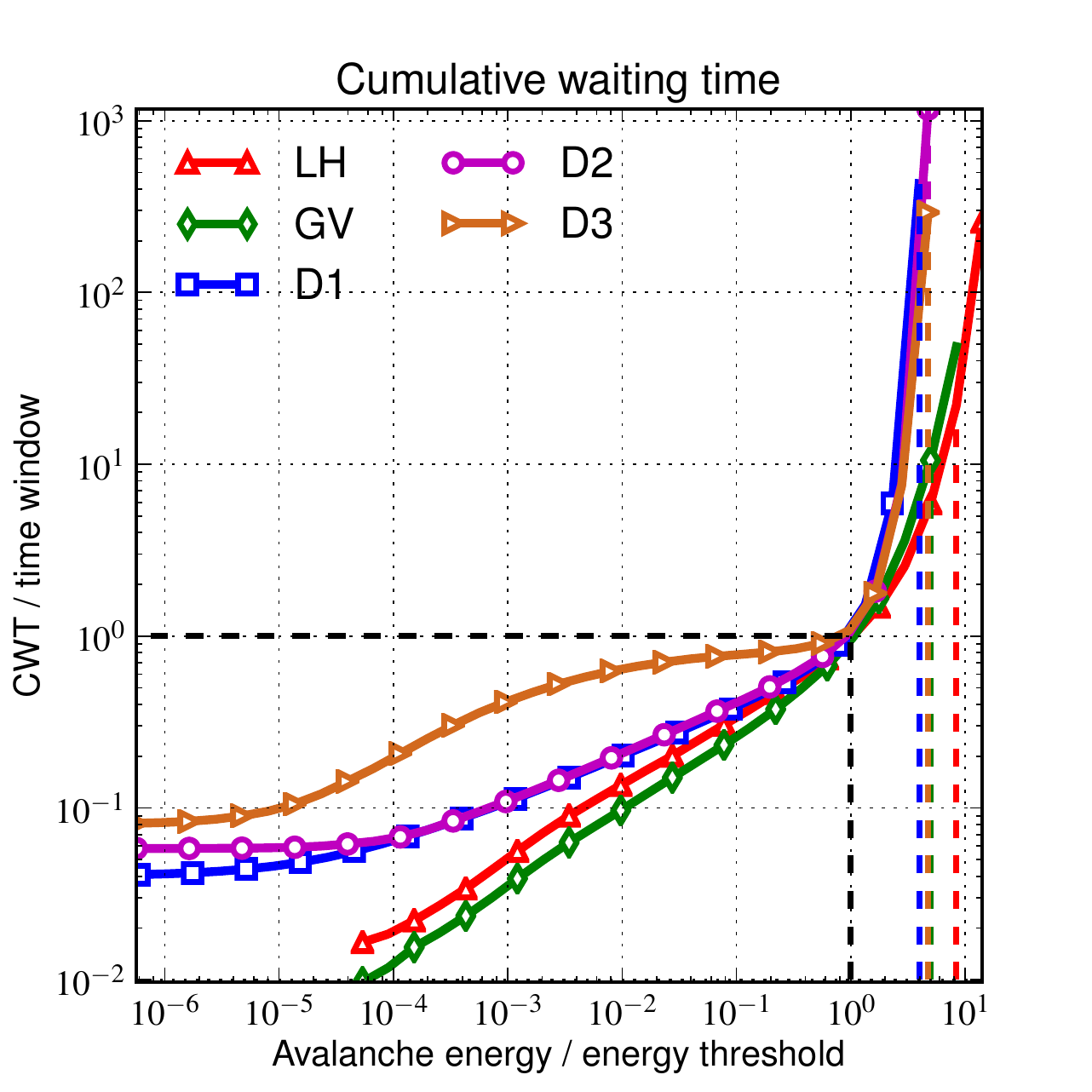}
  \caption{Cumulative waiting time against avalanche energy (see
    text). Axis are normalized to the time window and energy threshold
  considered, which can vary from one model to one another.}
  \label{fig:time_properties}
\end{figure}

\subsection{Prediction Time Windows}
\label{sec:pred-time-wind}

Predicting the occurrence of an event requires the definition of the
time window on which the prediction is given. The largest events --
which we want to predict -- are rare but obey specific statistical
rules which depend on the avalanche model considered. We define a cumulative
waiting time $\tau_{\rm CWT}(E)$ defined by the averaged waiting time
between avalanches of energy higher than $E$. We display the
cumulative waiting times as a function of the corresponding energy $E$
for the model we considered in 
fig. \ref{fig:time_properties}. \AStwo{The D models have an almost-constant
  waiting-time distribution for the population of small avalanches
  \citep[see][]{Strugarek:2014kj}. Then, the models exhibit of
a power-law part (except for model D3) followed by an exponential part for the highest energies.
This change reflects the energy limit the avalanches can
access due to the finite size of the lattices.}

\AStwo{For each model, we identify the time window $\tau_{w}$ and its
  corresponding energy $E_{0}$ at which the slope of the cumulative
 waiting times changes (these quantities are model-dependent and are
 used to normalize the axes in figure \ref{fig:time_properties}). We are interested in the
 capability of the models to predict the largest avalanches, hence we
 consider here avalanches of energy higher than $E_{0}$. We use in the
following the time window $\tau_{w}$ to asses the predictive
capabilities of the models. Because of the known statistical distribution of
avalanches, most of the runs carried over $\tau_{w}$ will trigger at
least an avalanche of energy $E_{0}$. In order to ensure that the
predictive skills of the models are not the result of a simple
statistical occurrence of an avalanche over $\tau_{w}$, we aim at
determining the predictability of avalanches of energy $E_{p}$ (and
higher) that statistically occur on time windows larger than $10\, \tau_{w}$. This will ensure that the
observed avalanche results from a particular lattice state and/or a
particular random number sequence (see section
\ref{sec:mean-prediction-from}), and is not the result of a
climatological forecast based on the overall properties of the
statistical SOC model
\citep[see also][]{Barnes:2008bw}.}


\subsection{Predictions from Stochastic Models}
\label{sec:mean-prediction-from}

\begin{figure}
  \centering
  \includegraphics[width=0.8\linewidth]{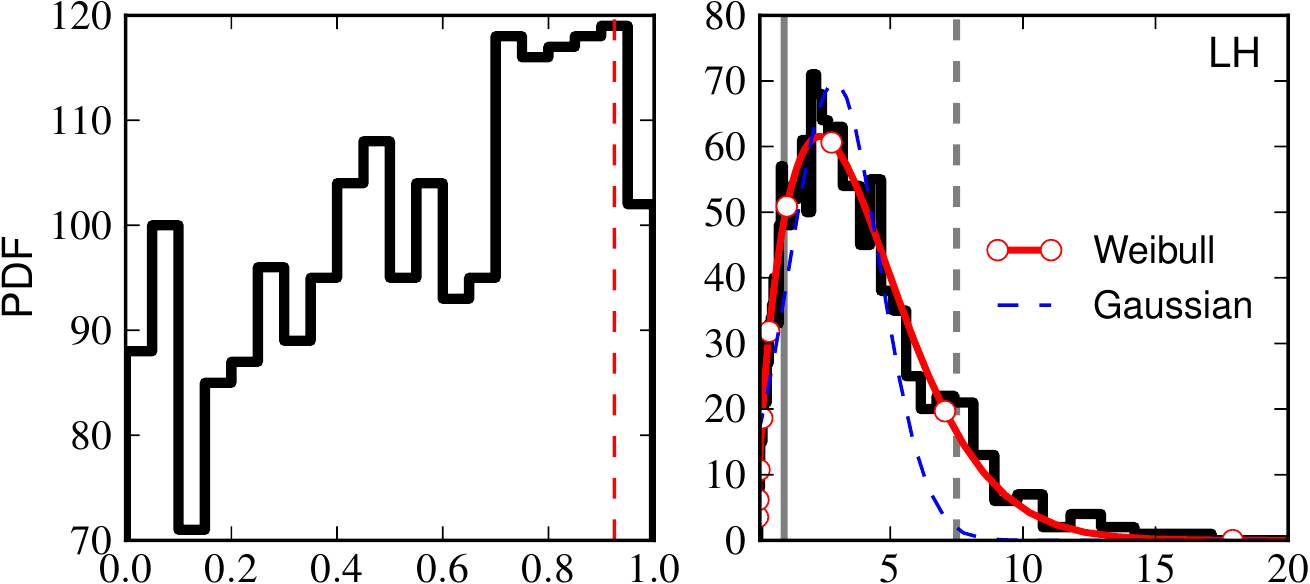}
  \includegraphics[width=0.8\linewidth]{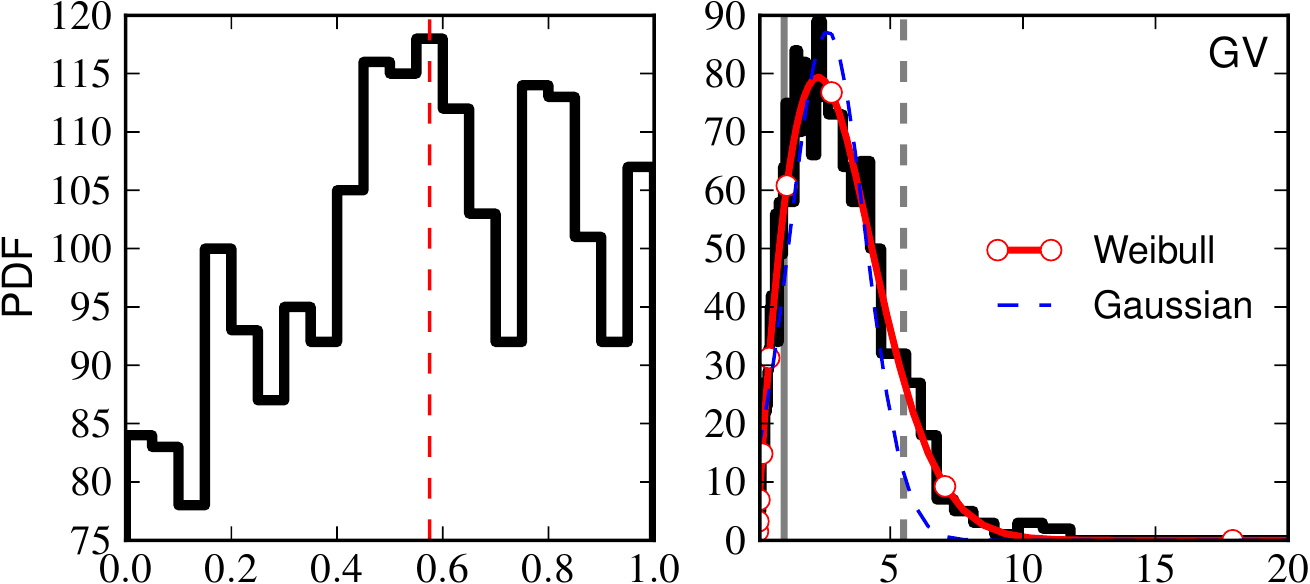}
  \includegraphics[width=0.8\linewidth]{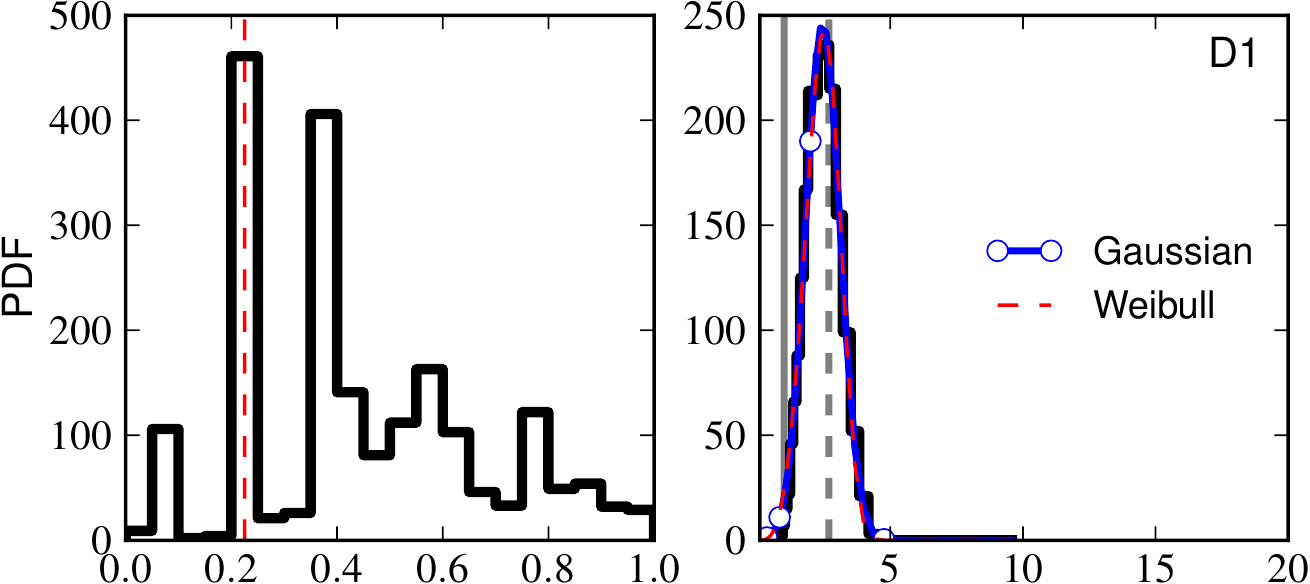}
  \includegraphics[width=0.8\linewidth]{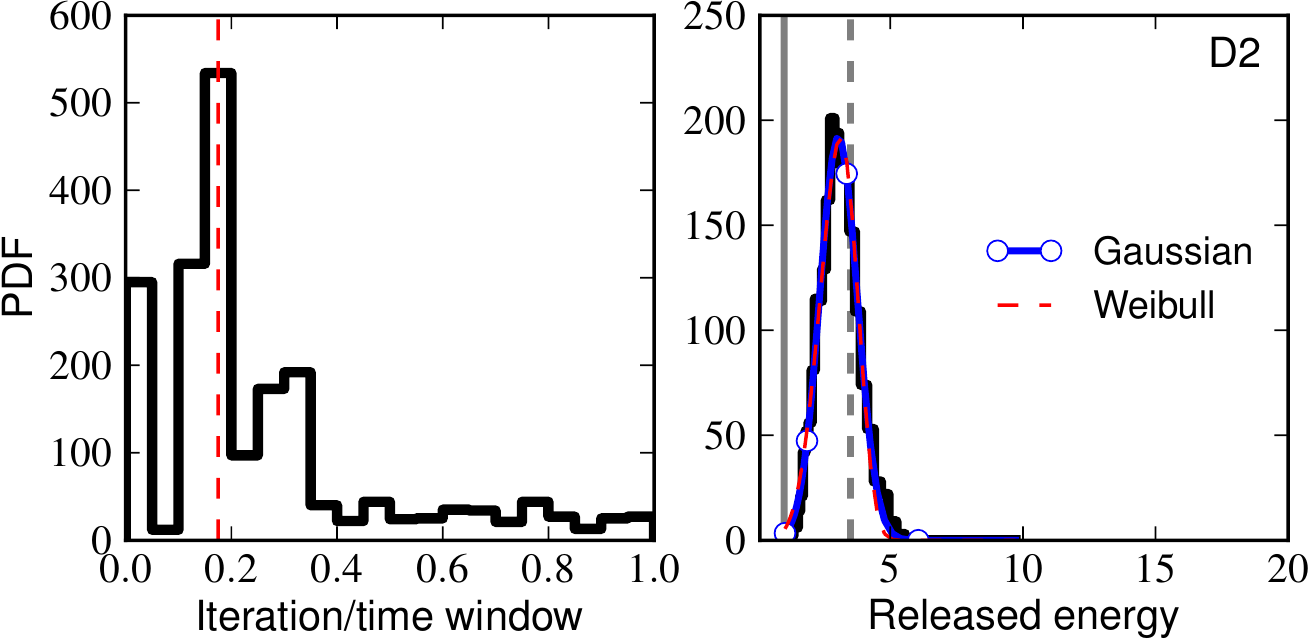}
  \caption{Largest avalanche mean prediction from a set of $2000$
    random number sequences \changed{using a common initial
      condition}. Left panels display the PDFs of the 
    predicted arrival time of the \changed{largest} avalanche, and right panels the
    PDFs of the predicted energy. \AStwo{The peak $\tau_{a}$} is identified
    by a vertical  red dashed line on left panels. The red and blue lines on the right panel
    label the Weibull and gaussian fits of the energy PDF. The
    solid and dashed vertical grey lines respectively label the
    normalization energy $E_{0}$ \changed{(the
      climatological forecast)} and the target energy $E_{\rm
      p}$. \AStwo{Results from D3 (not shown here) closely resemble D2.}
  }
  \label{fig:pred_example}
\end{figure}

Avalanche models always include a stochastic component, the
realization of which varies from one
run to one another. A prediction from a SOC-type model shall be composed of
a sufficient number of runs varying the stochastic component of the
model. Bearing in mind the analogy with solar flares prediction, we
want to predict what will be the energy of the largest avalanche in
the next $\tau_{w}$, and the time $\tau_{A}$ at which it will
occur. For one \changed{fixed} initial
condition, we use $2000$ different random number sequences and store
for each of them the starting time and release of energy of the
largest avalanche \changed{over the time window}. The resulting PDFs are
shown in fig. \ref{fig:pred_example} for the models \AStwo{LH, GV, D1
  and D2
\changed{for a representative initial condition of each
  model} (model D3 gives very similar results to D2)}. 

\changed{The left panels display the PDF of the occurrence time of the
  largest avalanche in the time window, and the right panels the PDF of the largest avalanche energy. 
In all models, we notice that the energy} is regularly distributed around a mean
value. While the D models are well fitted by Gaussian (blue lines), the first two
models have non-Gaussian tails. We chose to fit those resulting PDFs with a
Weibull function 
which give satisfying fits (red
lines). The best of the two fits is shown with a solid bulleted
line and the other is shown in dashed thin line, for reference. The
mean predicted value from the model is defined  
as the peak location of the best of the two fits. The fact that a mean can always be defined
confirms the intuition that the stress 
pattern embedded in an instantaneous SOC state indeed determines the
shape and size of upcoming avalanches, regardless of the stochastic
component of the model. The departure of LH and GV models from a
Gaussian behaviour nonetheless reveal an interesting property, which
shall be confirmed in the following. We observe that the PDF is heavy
tailed on the high energy side compared to a classical Gaussian
distribution. On the one hand, the stochastic process tend to give the
climatological prediction of the model ($E_{0}$, vertical solid grey line in fig.
\ref{fig:pred_example}). On the other hand, the \changed{initial} stress pattern allows
\changed{(or not)} for very large avalanches to be triggered ($E_{\rm p}$, vertical dashed grey
line). The heavy-tail part of the 
PDF directly derives from this latter property \changed{and is
  obtained only for a few favorable random number sequences},
while the stochastic process is strong enough to severely alter the mean prediction towards the
climatological forecast (vertical solid gray line). \changed{We
  clearly see that this property of the stochastic process is very
  strong in the LH and GV, and significantly weaker in models D. Hence, the
  Gaussian shape obtained in the D models simply indicative of marginal
  dependency upon the random number sequence: very large avalanche are
  triggered only when the adequate stress pattern exists in the lattice.} 

The left panels reveal another
fundamental difference -- in the context of predictability -- between
the models \changed{we considered}. The LH and GV models show PDFs of
$\tau_{A}$ spanning the whole time window\changed{, which means that
  the stochastic process determines when the largest avalanche will
  occur in the next time window. Hence, a} mean $\tau_{A}$ is
generally hard to define, although \changed{with some particular
  initial condition a slight peak value can be observed (not shown
  here).} Conversely, the 
D models show PDFs peaked for a few values of
$\tau_{A}$ which results in a very good confidence in the predicted time
for the maximal avalanche. We note nonetheless that the \Paulchanged{multiplicity
of peaks generally exhibited by the} D1 model
leads to a small but
significant uncertainty on the occurrence time of the avalanche
considered. \AStwo{Models D2 and D3 are the only models from which one can confidently
\changed{predict} an occurrence time.} 

The profound difference between the D models and the others is
naturally explained by the deterministic character \Paulchanged{of their driving process.
Being deterministic,} the driver dictates
unambiguously the next avalanching node which will be marginally
affected by the choice of random number sequence (we recall here that
\AStwo{in models D2 and D3}, the stochastic process is
only acting during avalanches and hence the driver completely
determines the next avalanching node). If this node is likely to trigger a large
avalanche, it will do so for most of the random number sequences. As a
consequence, the PDF of $\tau_{A}$ for the D models will always be peaked. 
\Paulchanged{For small avalanches the stochastic elements embodied in the D models
will significant impact the unfolding of a given avalanche, but for large avalanches,
where many hundreds of nodes are involved -- many avalanching repeatedly in
the course of the same avalanche --, these stochastic fluctuations
will tend to ``even out'' and have a limited impact on global avalanching
characteristic, including the amount of released energy.}

\subsection{Predictive Skills}
\label{sec:prediction-skills}

\begin{figure}
  \centering
  \includegraphics[width=0.75\linewidth]{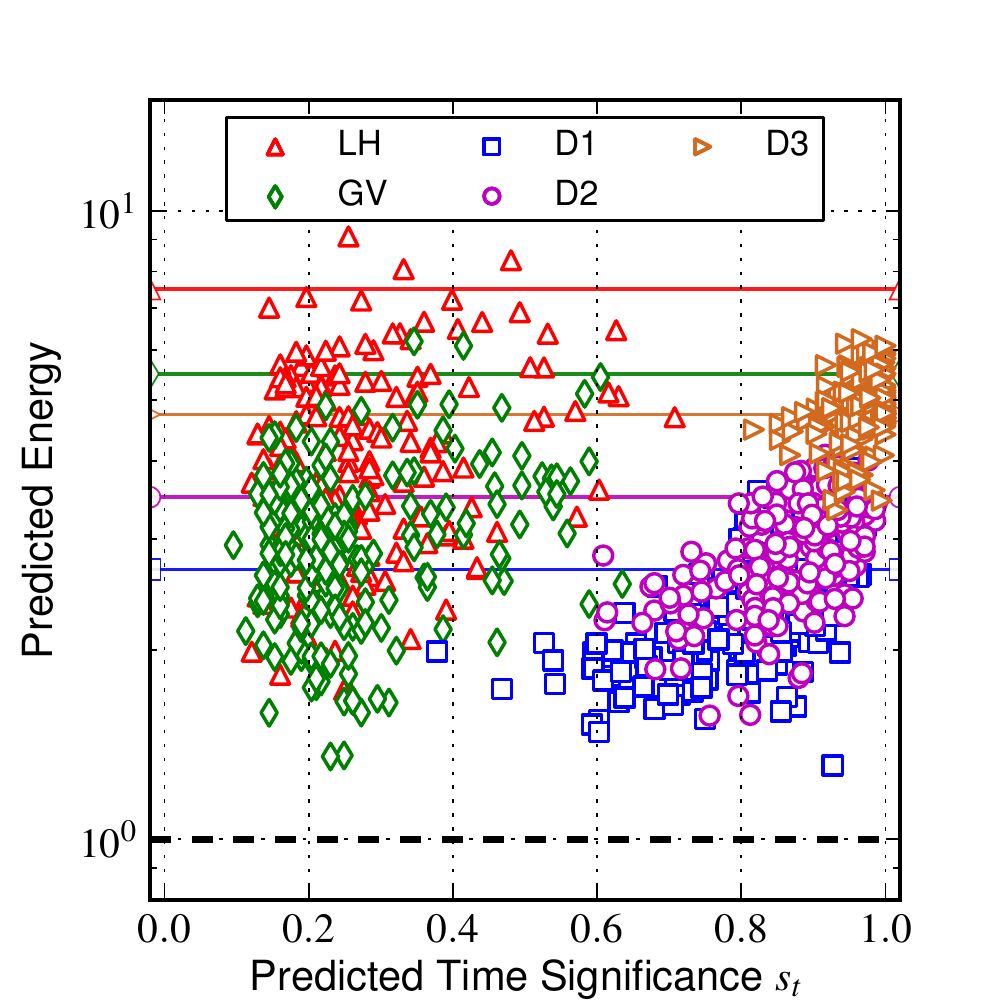}
  \caption{\changed{Predictive capabilities of the models for $200$ independent
    events. 
    The predicted energy (defined as
    the peak of the best fit of the PDF in the right panels of
    fig. \ref{fig:pred_example}) is shown against the predicted time
    significance (see text). The closer the points are from unity, the
    higher the predictive capability. The dashed horizontal
    line corresponds to the climatological forecast of the model, the
    thin light lines correspond to the targeted energy $E_{p}$ for event
    prediction in each model.}}
  \label{fig:combine_preds}
\end{figure}

We performed the statistical analysis described in
section \ref{sec:mean-prediction-from} for $200$ independent
\changed{initial conditions -- and associated large events --}
for each of the models. For each event, we define the predicted energy by the
mean of the best fit to the PDF of the predicted
avalanche energy. We also define the predicted time significance
$s_{t}$ by estimating the significance of the largest peak in the
$\tau_{A}$ distribution. A significance of $1$ corresponds to one
and only one peak at one particular time, while a significance of $0$
corresponds to a purely flat PDF (see the top two left panels of fig. \ref{fig:pred_example}).
We display the results of the $200$ cases for the \AStwo{five} models
in fig. \ref{fig:combine_preds}. The horizontal solid lines correspond to
the values of $E_{\rm p}$, and the dashed line
to the energy threshold. The intuition we got from fig.
\ref{fig:pred_example} is confirmed: the arrival time prediction is
very good for D models ($s_{t} \gtrsim 0.5$), while LH and GV models
rank poorly for almost all events ($s_{t} \lesssim 0.6$). The energy predicted
by the LH and GV models lies in between the ``climatological forecast''
($E_{0}$, dashed line) of the model and the target energy -- which
they almost never predict correctly. This implies that they are
unable to predict reliably the very large avalanches. \AStwo{Conversely,
large avalanches occurring in D models are generally recovered
regardless the particular random sequence. Model D1 exhibits intermediate
behavior with instances of both good and bad prediction, and model D3
performs extremely well, with $s_{t} > 0.8$ for all the events
considered here.}  

\begin{figure}
  \centering
  \includegraphics[width=\linewidth]{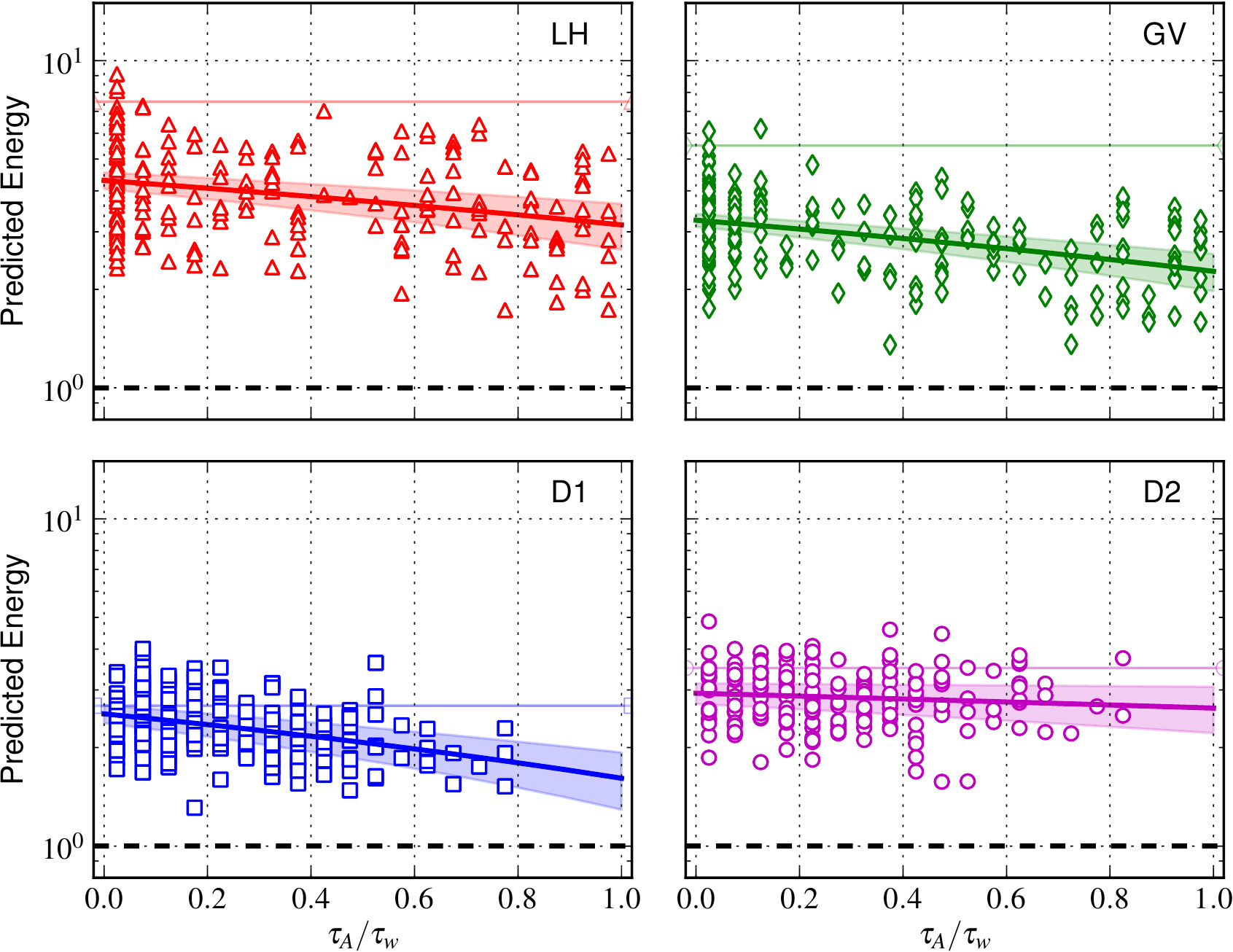}
  \caption{\changed{The same data than in fig. \ref{fig:combine_preds}
      is plotted against the occurrence time of the predicted
      avalanche, normalized to the prediction time
      window. The thick lines represent the best fit of the data
      (eqs. \ref{eq:fits_lin}-\ref{eq:fits_lin_end}), and the
      corresponding shaded areas the estimated error of the fit.}}
  \label{fig:combine_preds2}
\end{figure}

Some events of the LH and GV models have good $s_{t}$
significance and the corresponding predicted energy is significantly larger
than the other cases. They can be explained by taking a close look at
fig. \ref{fig:combine_preds2} which displays the same results as on
fig. \ref{fig:combine_preds} but against the largest avalanche occurrence time (normalized
to the time window) rather than $s_{t}$. The successfully predicted
events of the LH and GV models
correspond to runs where the large avalanches took place at the very
beginning of the runs. Hence, they occurred in runs where almost
no random numbers were involved, which is why they have very good
predictive capability. We observe that the later the large avalanche
occurs in the run, the lesser LH, GV and D1 models are able to predict an
avalanche different from the climatological forecast (dashed
line). \AStwo{The D2 model is, in contrast, predicting large
avalanches fairly accurately regardless of 
the occurrence time in the prediction time window, again proving
its high predictive capabilities. Finally, the D3 model (not shown
here) exhibit almost no dependency of the event occurrence upon the
time window. The dispersion of the predicted energies
is nevertheless comparable in all the D models. To confirm this
interpretation, we further fit a
linear relation between $E/E_{0}$ and $\tau_{A}/\tau_{w}$ for each of
the models
(thick lines in Fig. \ref{fig:combine_preds2}). We obtain, with energies normalized
to $E_{0}$ for each model:}

\begin{eqnarray}
  \label{eq:fits_lin}
  E_{\rm LH}/E_{0} &=& 4.30 - 1.16\, \frac{\tau_{A}}{\tau_{w}}\, , \\
  E_{\rm GV}/E_{0} &=& 3.25 - 0.98\, \frac{\tau_{A}}{\tau_{w}}\, , \\
  E_{\rm D1}/E_{0} &=& 2.54 - 0.92\, \frac{\tau_{A}}{\tau_{w}}\, , \\
  E_{\rm D2}/E_{0} &=& 2.93 - 0.29\, \frac{\tau_{A}}{\tau_{w}}\, , \\
  E_{\rm D3}/E_{0} &=& 4.8 \, .
  \label{eq:fits_lin_end}
\end{eqnarray}

\AStwo{Again, only models D2 and D3 exhibit a small dependency of the predicted
energy on the avalanche occurrence time $\tau_{A}$. The other models
are clearly too sensitive to this parameter to be considered for
reliable forecasts.}



\section{Conclusions}
\label{sec:conclusions}

In this paper, we have assessed the predictive capabilities of
\changed{a representative set of } avalanches models for solar flares.
\changed{Starting from the reference model of Lu \& Hamilton, we
  modified the type of stochasticity embedded in the avalanche model
  and subsequently characterized their predictive capabilities.} We
focused our study on the
prediction of large events, which are the rarest and presumably the
hardest and most important to predict. We showed that only the purely
deterministically driven model is able to predict a large event
reliably, while the classical Lu \& 
Hamilton model generally fails. We recall that the
deterministically driven models show a deficit in term of small
avalanches (Fig. \ref{fig:mod_properties}) and hence depart from the
classical SOC state. However, this property does not modify the
predictive capabilities of the model, which were assessed for the
larger, rarer avalanches, traditionally the hardest to predict. 

Avalanche models always include the two conflicting aspects (in terms of
predictive capabilities) of stochasticity and long-term
correlations. By exploring the physical interpretation of avalanche
models in the context of solar flares, we modified the stochastic process location in
the model which lead to the development of a deterministically-driven model
\citep{Strugarek:2014kj}. We empirically demonstrated that this model
possess the required properties to 
be used as a predictive tool: it is able to unambiguously predict the
large avalanche occurrences over a given time window, well above the
``climatological forecast''. We were able to demonstrate that the
deterministically driven model has very little bias with the event
occurrence time over a selected time-window. All the other models we considered are
significantly affected by \changed{they stochastic component} 
which makes them impossible to use for any practical prediction of large
events. Computationally, avalanche models also have the significant
advantage of being extremely inexpensive to run (this naturally
results from their low dimensionality and their simplicity). Those
\changed{unexpected} properties promote a further investigation for the development of a
near-real time prediction tool of large solar flares based on
deterministically driven avalanche models. \changed{We envision this a a
two-steps process.}

\AStwo{First, as noted already 
  the deterministically driven models considered here produce either power-law
  exponents (e.g., for the avalanche/flare energy distribution
  function) that are significantly lower than the real solar
  flare distribution exponents, or a small excess of large
  events. These different
    deterministically-driven models all possess good predictive
    capabilities for large events, which is probably one of their
    robust features. 
    As a result, variations of the D model could be explored \citep[e.g., in the
  spirit of][]{Strugarek:2014kj} to produce a model combining good
  predictive capabilities with statistical properties closer
  to the real solar flare data.}

\begin{figure}
  \centering
  \includegraphics[width=\linewidth]{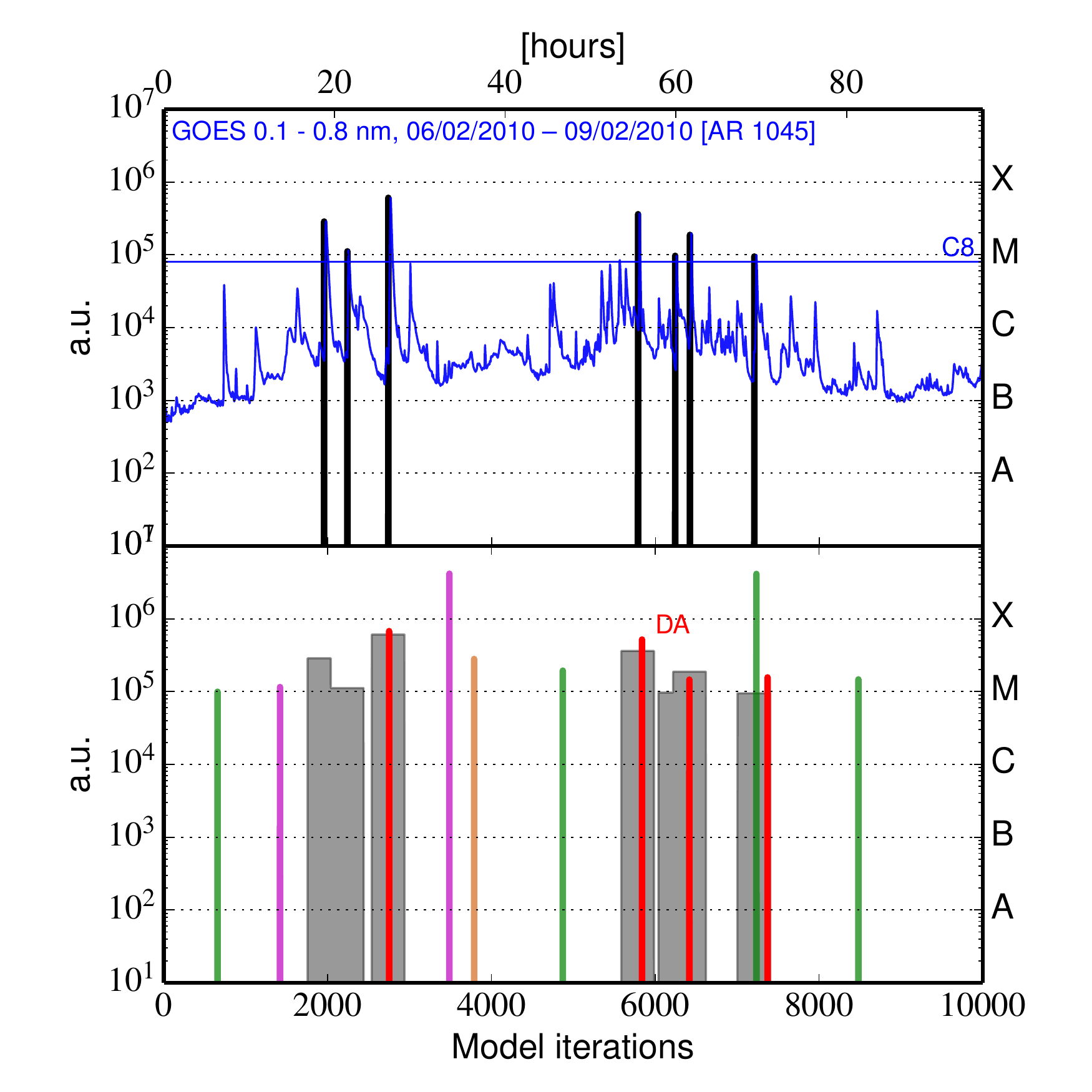}
  \caption{\ASthree{Example of a run using data assimilation for the
      GOES flux in the $1$--$8$ $\dot{A}$ range during the flaring
      events of the active region 1045 between February 6 and February
      9, 2010. The top panel show the GOES
      flux (blue) and the processed signal
      \citep[see][]{Aschwanden:2012ft} (black) for flares of class C8
      and above. The bottom panel shows the GOES signal (grey boxes)
      used in the data assimilation run. The assimilated sequence (DA) is
      shown in red along with three random realizations of the model in
      orange, green and magenta.}}
  \label{fig:ex_DA}
\end{figure}

\changed{The second step consists in the coupling of}
 data assimilation techniques \citep[see, e.g.][]{Belanger:2007ey}  to the \changed{chosen}
deterministically-driven model,
and use observed times series \ASthree{(such as the Geostationary Operational
Environmental Satellite -- GOES X-ray time series)} to
assess quantitatively the predictive skills of the model
\citep{Barnes:2008bw,Bloomfield:2012dj}. Our approach is complementary
to the approach of \citet{Wheatland:2005kf} who also used the
statistical properties of solar flare to predict the whole-Sun large events
occurrence. In our case, the avalanche model can be viewed as a
representation of one particular active region from which we will
assimilate data. \ASthree{The use of selected data
assimilation techniques will allow us to automatically adapt the driver and/or
lattice state of the model, so that the output of the model matches the
observed data. We show in figure \ref{fig:ex_DA} an example of such data
assimilation run using a simulated annealing method. The GOES flux
(blue line in top panel) is transformed
following to the method described in \citet{Aschwanden:2012ft} into a
distribution of delta-functions (black lines) filtered for flares of class C8
and above. We make the conversion of the GOES time sequence in terms
of avalanche energy and model iterations using the typical waiting
times of flares above C8 in model D3. In the bottom panel, we show
three random realizations of model D3 (sets of light colored vertical lines) and one
realization of model D3 using data assimilation (DA, red peaks) with the
observed GOES time series (gray boxes). Here the
assimilation technique succeeds in capturing the two clusters of
flares exceeding C8 at $\simeq 25$ and $\simeq 60$\, hr, without
generating spurious ($\geq$ C8) flares before, after, or in between. The
energy levels of the four reproduced flares also match very well with observations. The
details of this assimilation technique will be
described in details in \citet{Strugarek:2014uv}. These preliminary
results suggest that the lattice configuration resulting from the data
assimilation run combined with the predictive capabilities of the
model we demonstrated in this work could be used to carry out
quantitative predictions of solar flares through direct simulation.} We
believe that such a model
could lead to significant improvements of the current predictions of
large (typically X-class) solar flares.

\begin{acks}
The author thank the anonymous referee for valuable comments. The
authors acknowledge stimulating discussions during
the ISSI Workshops on Turbulence and Self-Organized Criticality
(2012-2013) held in Bern
(Switzerland); and during the ``Festival de th\'eorie'' (2013) held in
Aix-en-Provence (France). This research has made use of \href{http://sunpy.org/}{SunPy}, an
open-source and free community-developed solar data analysis package
written in Python \citep{Mumford:2013to}. We also acknowledge support from
the Natural Sciences and Engineering Research Council of Canada. AS acknowledges finanical support from CNES via SolarOrbiter grant.
\end{acks}

\bibliographystyle{spr-mp-sola}

\end{article}

\end{document}